\documentstyle[prb,aps,psfig]{revtex}

\begin{document}

\title{Time Dependent Theory for Random Lasers}

\author{Xunya Jiang, and C.~M.~Soukoulis}

\address{ Ames Laboratory-USDOE and Department of Physics and Astronomy,\\
Iowa State University, Ames, IA 50011\\}

\author{\parbox[t]{5.5in}
{\small
A model to simulate the
phenomenon of random lasing is presented. It couples  Maxwell's
equations with the
rate equations of electronic population in a disordered  system. Finite
difference time domain methods are used to obtain the field pattern and the
spectra of localized lasing modes inside the system. A critical pumping
rate $P_{r}^{c}$ exists for the appearance of the lasing peaks. 
The number of lasing modes increase
with the pumping rate and the length of the system. There is a lasing mode
repulsion.  This property leads to a
saturation of the number of modes for a given size system and a relation
between the localization length $\xi$ and average mode length $L_m$.\\ \\
PACS~numbers:~42.25.Bs, 72.55.Jr, 72.15.Rn, 05.40.-a}}

\maketitle

\newpage


The interplay of localization and amplification is an old and interesting
topic in physics research\cite{letok}. With promising properties, mirror-less
random
laser systems are widely studied\cite{expt,lawandy,zyu,john,li}  both
experimentally and theoretically.
Recently new observations \cite{expt} of laser-like emission were
reported and showed new interesting
properties of amplifying media with strong randomness. First, sharp
lasing peaks appear when the gain or the length of the system is
over a well defined threshold value. Although a drastic spectral
narrowing has been previously observed \cite{lawandy}, discrete lasing
modes were missing.  Second, more peaks appear
when the gain  or the system size further increases  over
the threshold. Third, the spectra of the lasing system is
direction dependent, not isotropic.
To fully explain such an unusual behavior of stimulated emission in random
systems with gain, we are in need of new theoretical ideas.

Theoretically, a lot of methods have been used to discuss the
properties of such
random lasing systems. Based on the time-dependent diffusion equation,
earlier work of Letokhov \cite{letok} predicted the possibility  of
lasing in
a random system and Zyuzin \cite{zyu} discussed the fluctuation
properties near the lasing threshold. Recently, John and Pang \cite{john}
studied the random lasing system by combining the electron
number equations of energy levels with the diffusion equation. Such a
consideration  predicted a reduction in the threshold gain for  a laser
action due
to  the increased optical path from diffusion. It also verified the narrowing
of the output spectrum when approaching the gain threshold. By using the
diffusion
approach is not possible to explain the lasing peaks observed in the
recent experiments\cite{expt} in both semiconductor powders and in
organic materials.
The diffusive description of
photon transport in gain media neglects the phase coherence of the
wave, so it gives limited information for the wave propagation in the
gain media.
Another approach which is based on the time-independent wave equations
for the random gain media can go beyond
the diffusive description \cite{li,pr,zhang,dual,jiang}. But as
was shown
recently \cite{jiang2}, the time-independent method
is only useful in determining  the
lasing threshold.
When the gain or the length of system is larger than the threshold value,
the time-independent description will give a totally unphysical
picture for such a system. To fully understand the random lasing system, we
have to deal with time-dependent wave equations in random systems.
   On the other hand,
the laser physics community \cite{lamb,lasers} has developed some
phenomenological
theories to deal with gain media which were overlooked by the researchers
working on random systems.

In this paper we introduce  a model by
combining these semi-classical laser theories with
Maxwell equations. By incorporating a well-established FDTD (
finite-difference time-domain) \cite{fdtd} method we
calculate the wave propagation in random media with gain. Because this
model couples electronic number equations at
different levels with field equations, the  amplification is  nonlinear
and saturated,
so stable state solutions can be obtained after a long relaxation time.
The advantages of this  FDTD model are obvious, since one
can  follow the evolution of the electric field and electron
numbers  inside the system. From the
field distribution inside the system, one can clearly distinguish the
localized modes from the extended ones. One can also examine the time
dependence of the electric field inside and just outside the system.
Then after Fourier  transformation, the emission spectra
 and the modes inside the system can be obtained.

Our system is essentially a one-dimensional simplification of the real
experiments \cite{expt,lawandy}. It consists of many dielectric
layers of real dielectric constant of fixed thickness, sandwiched
between two surfaces, with the spacing between the dielectric layers
filled with gain media (such as the solution of dye molecules). The distance
between the neighboring dielectric layers is assumed to be a random variable.
The overall length of the system is $L$.

Our results
can be summarized as follows: ($i$) As expected for periodic and $short$
($L<\xi$, $\xi$ is the localization length) random system, an
extended mode dominates the field and the spectra.      ($ii$) For either
strong
disorder or the long ($L \gg \xi$) system, we  obtain a
low threshold value for lasing. By increasing the length or the gain
(higher gain can be achieved by increasing the pumping intensity) more
peaks appear in the spectra. By examining the field distribution inside the
system, one can clearly see that these lasing peaks are coming from
localized modes. ($iii$) When the gain or the pumping intensity increases
even
further, the number of lasing modes do not increase further, but saturate
to a constant value, which is proportional to the length of system for a
given randomness. And ($i$v) the emission spectra are not same for
different output directions which show that the emission is not
isotropic. These findings are in agreement with recent experiments \cite{expt}
and also make new predictions.

The binary layers of the system are made of dielectric materials with
dielectric constant of $\varepsilon_1 =\varepsilon_0$
and $\varepsilon_2= 4 \times \varepsilon_0$ respectively. The thickness of
the first layer, which simulates the gain medium, is  a random
variable $a_n = a_0 (1+W\gamma)$ where $a_0=300$nm,  $W$ is the
strength of randomness and $\gamma$
is a random value in the range [-0.5, 0.5]. The thickness of second
layer, which simulates the scatterers, is a constant $b=180$nm.
 In the first layer, there
is a
 four-level electronic material mixed inside.  An
external mechanism pumps electrons
from ground level ($N_0$) to third level ($N_3$) at certain pumping rate
$P_r$, which is proportional to the  pumping intensity in
experiments. After a  short lifetime $\tau_{32}$,
 electrons can non-radiative transfer to the second level ($N_2$). The
second level ($N_2$) and the first level ($N_1$) are called the upper and the
lower  lasing levels. Electrons can be transfered from the upper to the lower
level by both spontaneous and stimulated emission.
At last, electrons can non-radiative transfer from the first level
($N_1$) back to the ground level ($N_0$). The lifetimes and
energies of upper and lower lasing levels
are $\tau_{21}$, $E_2$ and  $\tau_{10}$, $E_1$
respectively. The center frequency of radiation is $\omega_a=
{(E_2 - E_1)}/{\hbar}$ which is chosen to be equal to  $2\pi\times6 \cdot
10^{14}$ $Hz$
($\lambda=499.7$ $nm$).
Based on real materials \cite{lasers}, the parameters $\tau_{32}$,
$\tau_{21}$ and $\tau_{10}$ are chosen to be $1\times10^{-13}$s,
$1\times 10^{-9}$s
and $1\times 10^{-11}$s. The total electron density $N_0^0=N_0+N_1+N_2+N_3$
and the pump rate $P_r$ are the controlled variables according to the
experiments \cite{expt}. 

The time-dependent Maxwell equations are given by
 $\bf{{\nabla} \times \bf{E}} = -{\partial \bf{B}}/{\partial t}$ and
 $\bf{{\nabla} \times \bf{H}} = \varepsilon {\partial \bf{E}}/{\partial
t} + {\partial \bf{P}}/{\partial t}$, where $\bf{B}=\mu\bf{H}$ and
$\bf{P}$ is the electric polarization density from which the
amplification or gain can be obtained.
 Following the single electron case, one can show  \cite{lasers} that
 the polarization density in the presence of an electric field obeys the
following equation of motion:

\begin{equation}
\frac{d^2 P(t)}{d t ^2}+\Delta\omega_a \frac{d P(t)}{dt} +{\omega_a^2}P(t)
=\frac{\gamma_r}{\gamma_c}\frac{e^2}{m} \Delta N(t) E(t)
\end{equation}
where $\Delta\omega_a = {1}/{\tau_{21}} + {2}/{T_2}$ is the full width at
half maximum linewidth of the atomic transition. $T_2$ is the mean time
between dephasing
events which is taken to be   $2.18 \times 10^{-14}$s,
$\Delta N(t)=N_1-N_2$ and
$\gamma_r={1}/{\tau_{21}}$ is the real decay rate of the second level and
$\gamma_c=\frac{e^2}{m}\frac{{\omega_a}^2}{6\pi\varepsilon_0c^3}$ is the
classical rate.  It is easy to derive \cite{lasers} from Eq. (1) that
the amplification line shape is Lorentzian and 
homogeneously
broadened.  Eq. (1) can be thought as a quantum mechanically correct
equation for the induced polarization density $P(t)$ in a real atomic system.

The  equations giving the number of electrons on every level can be
expressed as follows:

\begin{eqnarray}
\frac{d N_3(t)}{d t} &=& P_r N_0(t) -\frac{N_3(t)}{\tau_{32}} \nonumber \\
\frac{d N_2(t)}{d t} &=& \frac{N_3(t)}{\tau_{32}} + \frac{1}{\hbar\omega_a}
E(t) \frac{d P(t)} {d t} - \frac{N_2(t)}{\tau_{21}} \nonumber \\
\\
\frac{d N_1(t)}{d t} &=& \frac{N_2(t)}{\tau_{21}} -\frac{1}{\hbar\omega_a}
E(t) \frac{d P(t)} {d t}  - \frac{N_1(t)}{\tau_{10}} \nonumber \\
\frac{d N_0(t)}{d t} &=& \frac{N_1(t)}{\tau_{10}} - P_r N_0(t)  \nonumber
\end{eqnarray}
where $\frac{1}{\hbar\omega}E(t) \frac{d P(t)} {d t}$ is the induced
radiation rate from level 2 to level 1 or excitation rate from level 1 to
level 2 depending on its sign.

To excite the system, we must introduce  sources into the system.
To simulate the real laser system, we introduce sources homogeneously
distributed in the system to simulate the spontaneous emission. We make
sure that the distance between the two sources $L_s$  is smaller than the
localization length $\xi$.  Each source generates waves of a
Lorentzian frequency
distribution centered around $\omega_a$, with its amplitude
depending on $N_2$. In real
lasers, the spontaneous
emission is the most fundamental noise \cite{lamb,lasers} but generally
submerged in other technical noises which are much larger. In our
system, the
simulated spontaneous emission is the $only$ noise present, and is
treated self-consistently. This is the reason for the small background in the
emission spectra shown below.

There are two leads, both with width of 3000 nm, at the
right and the left sides of the system and at  the
end of the leads we use the Liao method \cite{fdtd} to impose an
absorbing-boundary conditions (ABC).
 In the FDTD calculation, discrete time step and space steps are chosen 
to be 
 $10^{-17}$s and $10^{-9}$m respectively. Based on the previous time steps
we can calculate the next time step ($n+1$ step) values. 
First we obtain the $n+1$  time step of the electric polarization 
density 
$P$ by using Eq. (1), then the  $n+1$ step of the electric and magnetic 
fields are obtained by Maxwell's equations and at last the  $n+1$ step 
of the electron numbers at each level are calculated by Eq. (2).
The initial state is that all
electrons are on the ground state, so there is no field,  no polarization
and
no spontaneous emission. Then the electrons are pumped and the system
begins to evolve according to equations.

   We have performed numerical simulations for periodic and
random systems.
First, for all the systems,  a well defined lasing threshold
exists. As expected, when the
randomness becomes stronger,
the threshold intensity decreases because localization effects make the
paths of waves  propagating inside the gain medium much longer.

For a periodic or $short$ ($L<\xi$) random system, 
generally only one mode dominates the whole system even if 
the gain increases far above the threshold. This is due to the fact that the
first mode can
extend in the whole system, and its strong electric field can force almost
all the electrons of the upper level $N_2$ to jump down to the
$N_1$ level quickly by stimulated emission.  This leaves very few
upper electrons for stimulated emission of the other modes. In other
words, all the other modes are
suppressed by the first lasing mode even though their threshold values are
only a little bit smaller
than the first one. This phenomenon also exists in common lasers
\cite{lamb}.

  For $long$ ($L>>\xi$) random systems, richer behavior is observed.
First we find that all the lasing modes are localized
and stable
around their localization centers after a long time. Each mode
has its own specific frequency and corresponds to a peak in the
 spectrum inside the system.
When the gain increases beyond the threshold, the electric  field pattern (see
Fig. 1a)   shows that more localized lasing modes appear in the system and
the spectrum inside the system (see Fig. 1b)
gives  more sharp peaks just as observed in the experiments \cite{expt}.
This is
clearly seen in Figs 1a and 1b for 
 a 80 cell random system, above
threshold.  In Figs. 1c and 1d  similar values are shown for the 160 cell
system.
Notice that both the number of localized modes (Fig. 1c) of the field, as
well the
number of lasing peaks (Fig. 1d) are larger now. The exact position of the
lasing
peaks depends on the random configuration.
 Notice that the lasing peaks are
much narrower than the experimental ones \cite{expt}. This is due to the 1d
nature of our model. In the present case only two escaping channels exist,
so it's more difficult for the wave to get out from the system which has a 
higher quality factor. When the gain is really big, we find  the number of lasing 
modes 
will not increase any more, so a saturated  number $N_m$ of lasing modes exists
for the
long random system. This is clearly seen in Fig. 2, where we plot the number of
modes $N_m$ vs the pumping rate $P_r$. In Fig. 3, we plot the spectral
intensity vs the wavelength for different input transitions (or
equivalently pumping rates). Notice  these results are in qualitative
agreement with the experimental results shown in Fig. 2 of the paper 
of~Cao~et.~al.~\cite{expt}.

  These multi-lasing peaks and the saturated-mode-number
phenomena are due to the  interplay between localization
and amplification.  Localization makes the lasing mode  strong
around its localization center and exponentially small away from its center
so that it only suppress the modes in this
area by reducing $N_2$.  When a
mode lases,
 only those modes which are $far$ $enough$ from this mode can
lase afterwards.  So  more than one mode can appear for a long system and
each mode seems to $repel$ each other.
Because every lasing mode dominates a certain area and is separated from
other modes, only
 a limited number of lasing modes can exist for a finite long system
even in the case of large amplification. We therefore expect that the number of
surviving lasing modes
$N_m$ should be
proportional to the length of the system $L$ when the amplification is very
large.  Since the "mode-repulsion" property
is coming from the localization of the modes we  expect that the
average
mode length $L_m=L/N_m$ should be proportional to localization length $\xi$ too.
In Fig.4, we plot $N_m$ vs the
length of the systems $L$ when we
increase the length from 80 cells to 320 cells and keep all other
parameters the same.
In Fig.4, we also plot the average mode length $L_m$ vs the localization
length
$\xi$ when  we change the random strength W for a 320 cell
system.  The
localization lengths are calculated using the transfer-matrix method by
averaging
10,000 random configurations. These results confirm that indeed
$N_m \propto L$ and $L_m \propto \xi$. It will be very interesting if
these  predictions can be checked experimentally.

  The emission spectra at the  right and left side of the
system  are quite different.
This can be explained from the field patterns shown in Fig. 1a and Fig. 1c.
Notice
the  localized modes are not similar at both sides of the system.  This is the
reason for  this difference in the output spectrum. In Fig. 1d we denote
with $l$
and $r$ the output modes from the left and the right side of our 1d system,
respectively. 
The non-isotropic output
spectra of real
$3D$ experiments
\cite{expt} might be explained by assuming that every localized mode
has  its intrinsic direction,
strength and position,  and the detected output spectra in
experiments at
different directions are the overlap of contributions from many modes. So
generally they should be different. 
Most of the modes are not able to escape in our model
and this is because of  the 1D localization effects and exchange of energy 
between modes.



In summary, by using a FDTD method we constructed a random four-level lasing 
model to study the interplay of localization and amplification.
Unlike the time-independent models, the present formulation
calculates the field
evolution beyond the threshold amplification. 
This model
allows us to obtain  the field pattern and spectra of
localized lasing modes inside the system. For  random systems, we can
explain the
 multi-peaks and the non-isotropic properties in
the emission spectra, seen experimentally. Our numerical results predict
the "mode-repulsion" property, the lasing-mode saturated number and average 
modelength. We also observed the exchange of energy between the localized 
modes which is much different from common lasers and this is essential for 
further research of mode competition and evolution in random laser. 
All of these properties are from the  interplay of the localization and
amplification where  new  physics phenomena can be found.

Ames Laboratory is operated for the U.S. Department of Energy by Iowa
State University under Contract No. W-7405-Eng-82. This work was
supported by the director for Energy Research, Office of Basic Energy
Sciences.

\begin{figure}
\psfig{figure=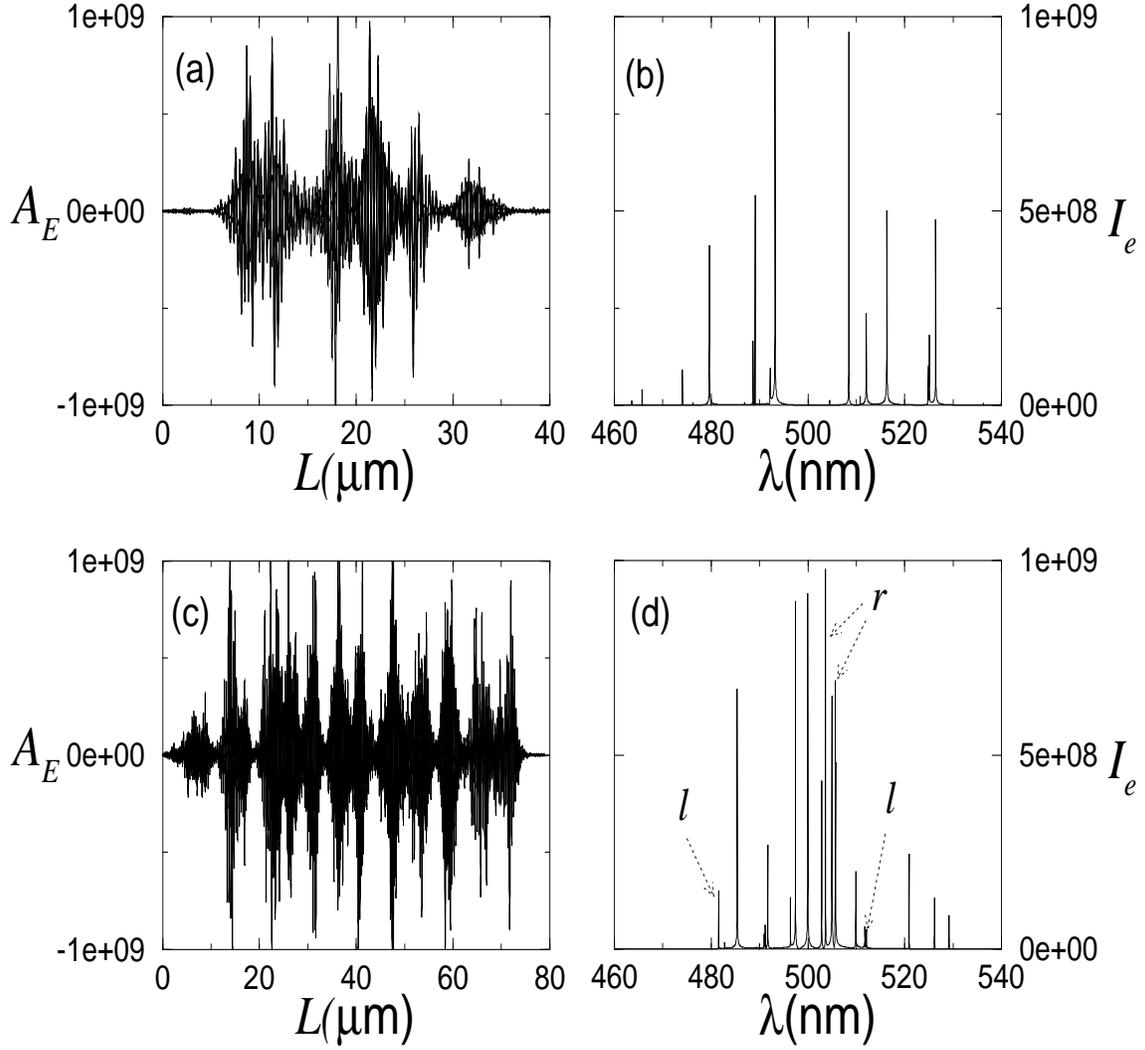,width=16cm,height=16cm,angle=270}
\caption{(a) The amplitude of the electric field $A_E$ vs the length of the
system and (b) the spectra intensity $I_e$ vs  the wavelength
for a 80 cell system with $W=1.2$, ${N_0}^0=5.5\times6.02\times10^{23}/m^3$
and $P_r=1\times10^{10}$$s^{-1}$.  In (c) and (d) $A_E$ and $I_e$ for a
160 cell system. }
\end{figure}

\begin{figure}
\psfig{figure=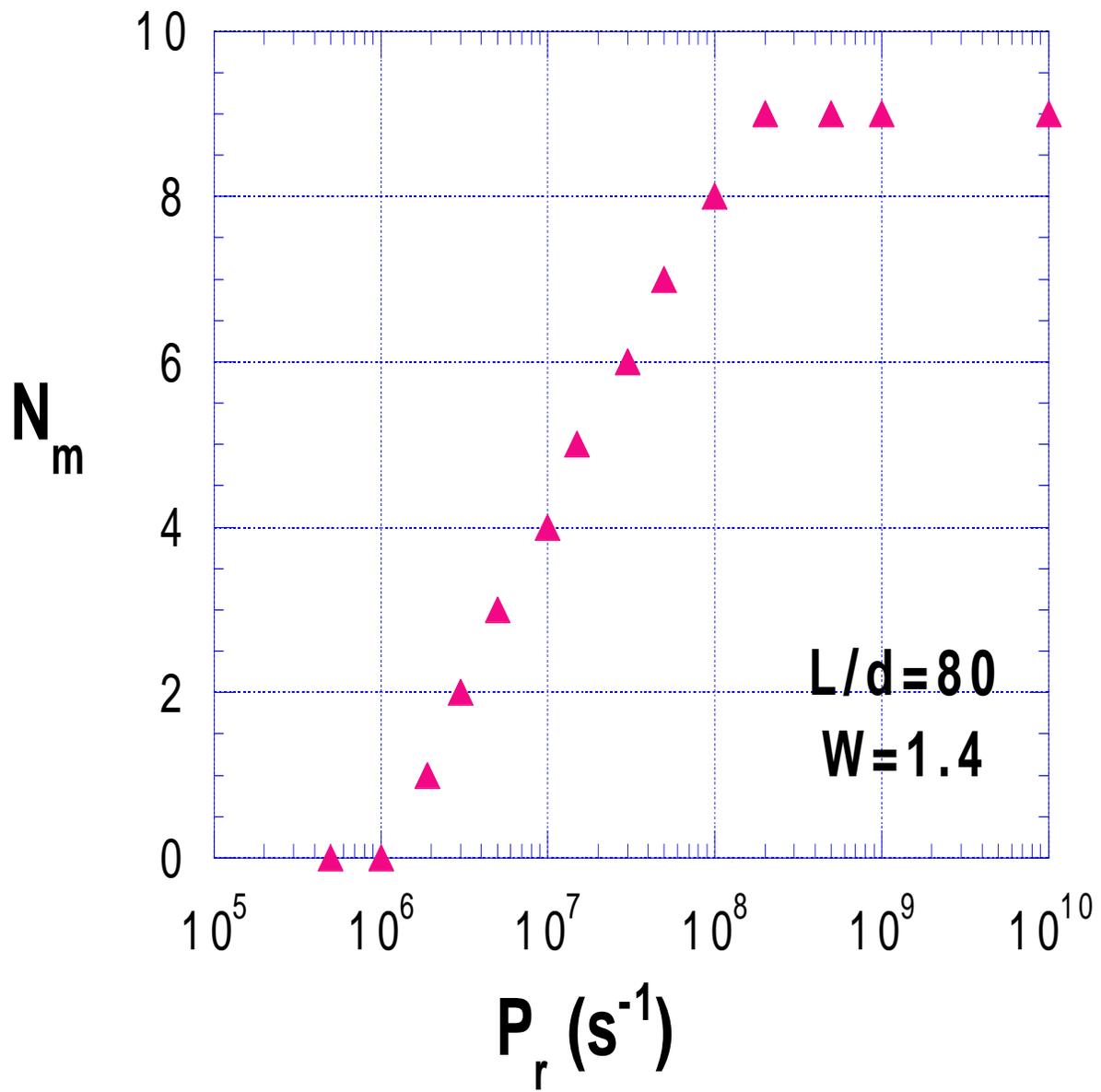,width=16cm,height=16cm,angle=0}
\caption{The number of modes $N_m$ versus the pumping rate $P_r$, for a 80 cell
system with W=1.4.  The critical pumping rate  $P_{r}^{c}$=$10^{6}s^{-1}$}.
\end{figure}

\begin{figure}
\psfig{figure=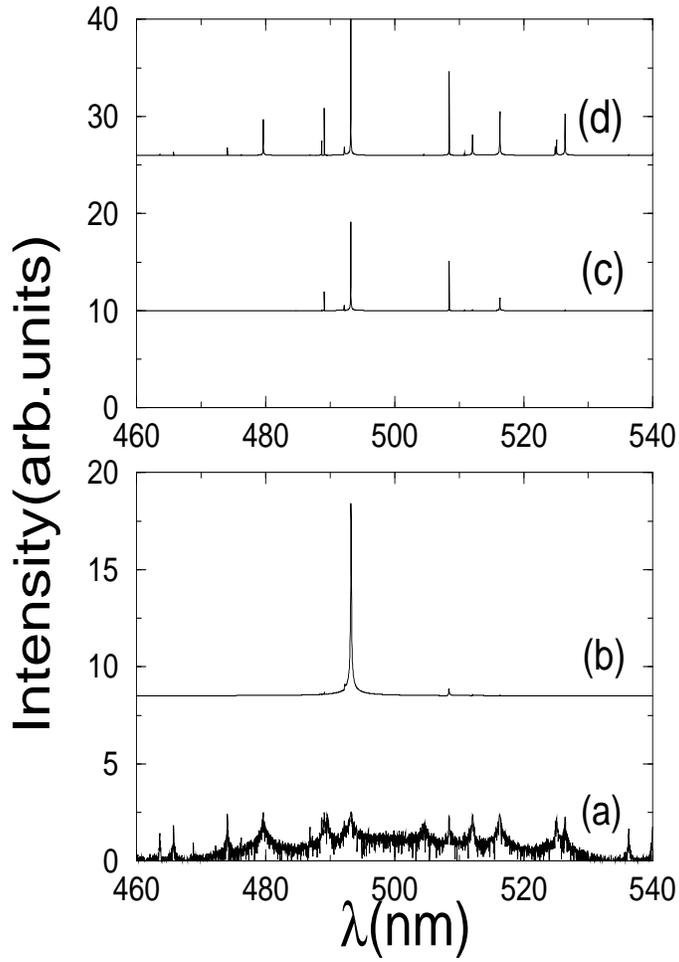,width=16cm,height=16cm,angle=270}
\caption{The spectral intensity vs the wavelength for the 80 cell system with
W=1.4 for different pumping rates $P_r$. $P_r$ in units of $s^{-1}$ is (a)
$10^{4}$, (b) $10^{6}$, (c) $10^{7}$ and (d) $10^{10}$}.
\end{figure}

\begin{figure}
\psfig{figure=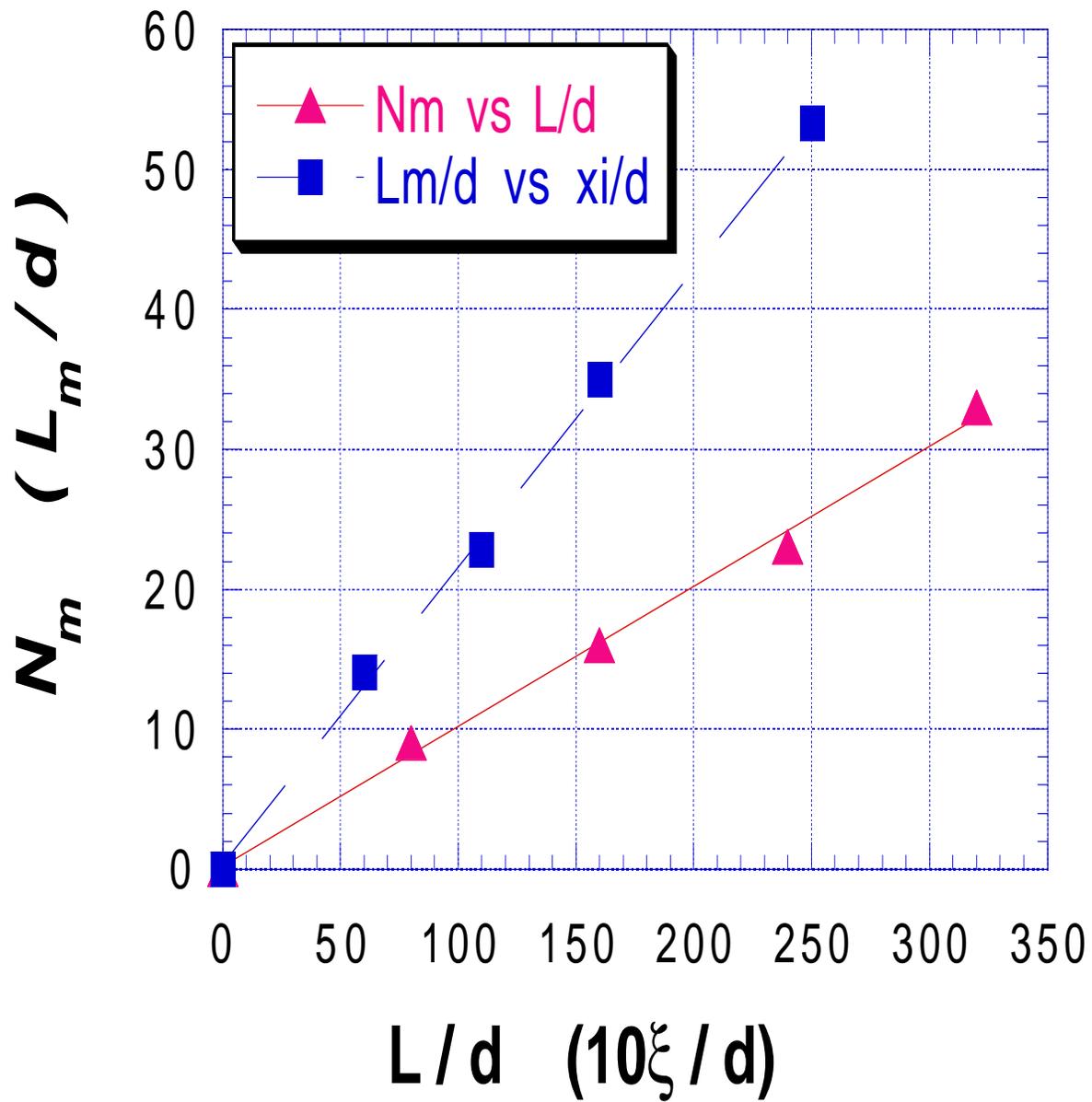,width=16cm,height=16cm,angle=0}
\caption{
The number of modes $N_m$ vs the length of the system $L/d$, where
d=$<a_n>$+b=480
nm is the size of the cell. Also the average mode length $L_m/d$ vs the
localization  length $10\times \xi/d$ for different disorder strength W for a 320 cell
system. The rest of the parameters are the same as the ones of Fig. 1.}
\end{figure}

\end{document}